\documentclass[12pt]{article}
\usepackage{a4wide}
\usepackage{epsfig}
\usepackage{amsmath}

\newlength{\absize}
\catcode`@=11
\def\citer{\@ifnextchar
[{\@tempswatrue\@citexr}{\@tempswafalse\@citexr[]}}

\def\@citexr[#1]#2{\if@filesw\immediate
  \write\@auxout{\string\citation{#2}}\fi
  \def\@citea{}\@cite{\@for\@citeb:=#2\do
    {\@citea\def\@citea{--\penalty\@m}\@ifundefined
       {b@\@citeb}{{\bf ?}\@warning
       {Citation `\@citeb' on page \thepage \space undefined}}%
\hbox{\csname b@\@citeb\endcsname}}}{#1}} \catcode`@=12

\begin{document}
  \thispagestyle{empty}
  \pagestyle{empty}
  \renewcommand{\thefootnote}{\fnsymbol{footnote}}
\newpage\normalsize
    \pagestyle{plain}
    \setlength{\baselineskip}{4ex}\par
    \setcounter{footnote}{0}
    \renewcommand{\thefootnote}{\arabic{footnote}}
\newcommand{\preprint}[1]{%
  \begin{flushright}
    \setlength{\baselineskip}{3ex} #1
  \end{flushright}}
\renewcommand{\title}[1]{%
  \begin{center}
    \LARGE #1
  \end{center}\par}
\renewcommand{\author}[1]{%
  \vspace{2ex}
  {\Large
   \begin{center}
     \setlength{\baselineskip}{3ex} #1 \par
   \end{center}}}
\renewcommand{\thanks}[1]{\footnote{#1}}

\begin{center}
{\large \bf Consistent Deformed Bosonic Algebra in Noncommutative
Quantum Mechanics}
\end{center}
\begin{center}
Jian-Zu Zhang
\end{center}
\begin{center}
Institute for Theoretical Physics, East China University of
Science and Technology, \\
Box 316, Shanghai 200237, P. R. China
\end{center}
\vspace{1cm}
\begin{abstract}
In two-dimensional noncommutive space for the case of both
position - position and momentum - momentum noncommuting, the
consistent deformed bosonic algebra at the non-perturbation level
described by the deformed annihilation and creation operators is
investigated. A general relation between noncommutative parameters
is fixed from the consistency of the deformed Heisenberg - Weyl
algebra with the deformed bosonic algebra. A Fock space is found,
in which all calculations can be similarly developed as if in
commutative space and all effects of spatial noncommutativity are
simply represented by parameters.
\end{abstract}

\clearpage
\section{Introduction}
\setcounter{equation}{0}
Physics in noncommutative space \citer{CDS,DN} has been
extensively investigated in literature. In the low energy sector
one expects that quantum mechanics in noncommutative space (NCQM)
may clarify some low energy phenomenological consequences, lead to
qualitative understanding of effects of spatial noncommutativity
and shed some light on the problem at the level of noncommutative
quantum field theory. In literature the perturbation aspects of
NCQM and its applications \citer{CST,Wu} have been studied in
detail. The perturbation approach is based on the Weyl - Moyal
correspondence \citer{AW,RS}, according to which the usual product
of functions should be replaced by the star-product. Because of
the exponential differential factor in the Weyl - Moyal product
the non-perturbation treatment is difficult. But non-perturbation
investigations may explore some essentially new features of NCQM.
A suitable example for non-perturbation investigations is a two
dimensional isotropic harmonic oscillator which is exactly
soluble, and fully explored in literature. In the first paper of
Ref.~\citer{JZZ04a} through the non-perturbation investigation of
this example it was clarified that the consistent ansatz of
commutation relations of phase space variables should
simultaneously include space-space noncommutativity and
momentum-momentum noncommutativity; The consistent deformed
bosonic algebra at the non-perturbation level was obtained and a
relation between noncommutative parameters was fixed. But this
example is special. In many systems, the potential can be modelled
by a harmonic oscillator through an expansion about its minimum.
But there are some potentials which are not the case. Since NCQM
is peculiar in many ways, it is necessary to clarify the situation
for general cases in detail.

In this paper we elucidate this topic for general cases in the
context of non-relativistic quantum mechanics. The point is how to
maintain Bose-Einstein statistics at the non-perturbation level
described by the deformed annihilation and creation operators in
noncommutative space when the state vector space of identical
bosons is constructed by generalizing one-particle quantum
mechanics. For this purpose, first we need to find the general
structure of the deformed annihilation and creation operators
which satisfy a complete and closed deformed bosonic algebra at
the non-perturbation level. We obtain the consistent deformed
bosonic algebra at the non-perturbation level. A relation between
noncommutative parameters is fixed from the consistency of the
deformed Heisenberg - Weyl algebra with the deformed bosonic
algebra. A construction of Fock space in general cases at the
non-perturbation level is complicated. We found a Fock space in
which all calculations can be similarly developed as if in
commutative space and all effects of spatial noncommutativity are
simply represented by parameters, not represented by
noncommutative operators.

In the following, in section 2 the background of the deformed
Heisenberg - Weyl algebra is reviewed. In section 3 the consistent
deformed bosonic algebra is investigated. In section 4 the
two-dimensional isotropic harmonic oscillator is revisited. Its
exact (non-perturbative) eigenvalues are obtained in a simple way
in commutative Fock space.

\section{The Deformed Heisenberg - Weyl Algebra}
\setcounter{equation}{0}
In this paragraph we review the necessary background first. The
starting point is the deformed Heisenberg - Weyl algebra. We
consider the case of both position - position noncommutativity
(space-time noncommutativity is not considered) and momentum -
momentum noncommutativity. In this case the consistent deformed
Heisenberg - Weyl algebra is \cite{JZZ04a}:
\begin{equation}
\label{Eq:xp}
[\hat x_{i},\hat x_{j}]=i\xi^2\theta\epsilon_{ij}, \qquad [\hat
p_{i},\hat p_{j}]=i\xi^2\eta\epsilon_{ij}, \qquad
[\hat x_{i},\hat p_{j}]=i\hbar\delta_{ij},\;(i,j=1,2),
\end{equation}
where $\theta$ and $\eta$ are constant parameters, independent of
the position and momentum. Here we consider the noncommutativity
of the intrinsic canonical momentum.
\footnote {\; The intrinsic noncommutativity of the canonical
momenta discussed here is essentially different from the
noncommutativity of the mechanical momenta of a particle in an
external magnetic field with a vector potential $A_i(x_j)$ in
commutative space. In the former case the difference between $\hat
p_i$ and $p_i$ must be extremely small, see footnote 3. In the
later case the mechanical momentum is
$$p_{mech,i}=\mu \dot x_i=p_i-\frac{q}{c}A_i,$$
where $p_i=-i\hbar \partial_i$ is the canonical momentum in
commutative space, satisfying $[p_i,p_j]=0$. The commutator
between $p_{mech,i}$ and $p_{mech,j}$ is
$$[p_{mech,i},p_{mech,j}]=-\frac{q}{c}
\left([p_i,A_j]+[A_i,p_j]\right)= i\frac{\hbar
q}{c}\left(\partial_i A_j-\partial_j A_i\right)=i\frac{\hbar
q}{c}\epsilon_{ij3}B_3.$$
Such a noncommutativity is determined by the external magnetic
field $\vec B$ which, unlike the noncommutative parameter $\eta$,
may be strong so that the difference between $p_{mech,i}$ and
$p_i$ may not be extremely small.}
It means that the parameter $\eta$, like the parameter $\theta$,
should be extremely small. This is guaranteed by a direct
proportionality provided by a constraint between them (See
Eq.~(\ref{Eq:c-2}) below). The $\epsilon_{ij}$ is a
two-dimensional antisymmetric unit tensor,
$\epsilon_{12}=-\epsilon_{21}=1,$ $\epsilon_{11}=\epsilon_{22}=0$.
In Eq.~(\ref{Eq:xp}) the scaling factor
$\xi=(1+\theta\eta/4\hbar^2)^{-1/2}$ is a dimensionless constant.

The deformed Heisenberg - Weyl algebra (\ref{Eq:xp}) can be
realized by undeformed phase space variables $x_{i}$ and $p_{i}$
as follows (henceforth summation convention is used)
\begin{equation}
\label{Eq:hat-x-p}
\hat x_{i}=\xi(x_{i}-\frac{1}{2\hbar}\theta\epsilon_{ij}p_{j}),
\quad
\hat p_{i}=\xi(p_{i}+\frac{1}{2\hbar}\eta\epsilon_{ij}x_{j}),
\end{equation}
where $x_{i}$ and $p_{i}$ satisfy the undeformed Heisenberg - Weyl
algebra
$[x_{i},x_{j}]=[p_{i},p_{j}]=0,\;
[x_{i},p_{j}]=i\hbar\delta_{ij}.$

It should be emphasized that for the case of both position -
position and momentum - momentum noncommuting the scaling factor
$\xi$ in Eqs.~(\ref{Eq:xp}) and (\ref{Eq:hat-x-p}) guarantees
consistency of the framework, and plays an essential role in
dynamics. One may argues that only three parameters $\hbar$,
$\theta$ and $\eta$ can appear in three commutators (\ref{Eq:xp}),
thus $\xi$ is an additional spurious parameter and can be set to
$1.$ If one re-scales $\hat x_{i}$ and $\hat p_{i}$ so that
$\xi=1$ in Eqs.~(\ref{Eq:xp}) and (\ref{Eq:hat-x-p}), it is easy
to check that Eq.~(\ref{Eq:hat-x-p}) leads to
$[\hat x_{i},\hat
p_{j}]=i\hbar\left(1+\theta\eta/4\hbar^2\right)\delta_{ij},$
thus the Heisenberg commutation relation cannot be maintained.

\section {The Consistent Deformed Bosonic Algebra}
\setcounter{equation}{0}
The investigation of the deformed bosonic algebra at the
non-perturbation level includes two aspects. The first aspect is
to find the general structures of the deformed annihilation and
creation operators which satisfy the complete and closed deformed
bosonic algebra at the non-perturbation level. Because there is a
new type of the deformed bosonic commutation relation which
correlates different degrees of freedom at the level of the
deformed annihilation and creation operators, the second aspect
is, by generalizing one - particle quantum mechanics to many
particle system, how to establish the Fock space of identical
bosons, see footnote 5 below.

We review the first respect \cite{JZZ04a}. In the context of
quantum mechanics the general representations of the deformed
annihilation and creation operators $\hat a_i$ and $\hat
a_i^\dagger$ at the non-perturbation level are represented by the
deformed phase space variables $\hat x_i$ and $\hat p_i$ as
follows:
\begin{equation}
\label{Eq:hat-a}
\hat a_i=c_1(\hat x_i+ic_2\hat p_i),\;
\hat a_i^\dagger=c_1(\hat x_i-ic_2\hat p_i),
\end{equation}
where $c_1$ and $c_2$ are real constants which can be fixed as
follows. Operators $\hat a_i$ and $\hat a_i^\dagger$ should
satisfy the bosonic commutation relations $[\hat a_1,\hat
a_1^\dagger]=[\hat a_2,\hat a_2^\dagger]=1$ (to keep the physical
meaning of $\hat a_i$ and $\hat a_i^\dagger$ at the
non-perturbation level). From this requirement and the deformed
Heisenberg - Weyl algebra (\ref{Eq:xp}) it follows that
\begin{equation}
\label{Eq:c1-c2}
c_1=\sqrt{1/2\hbar c_2}.
\end{equation}
Following the standard procedure in quantum mechanics, starting
from a system with one particle, at the level of the annihilation
and creation operators the state vector space of a many-particle
system can be constructed by generalizing one - particle
formalism. Then Bose - Einstein statistics for an identical -
boson system can be developed in the standard way. Bose - Einstein
statistics should be maintained at the non-perturbation level
described by $\hat a_i$, thus operators $\hat a_i$ and $\hat a_j$
should be commuting: $[\hat a_i,\hat a_j]=0$. From this equation
and the deformed Heisenberg - Weyl algebra (\ref{Eq:xp}) it
follows that
\begin{equation}
\label{Eq:B-E}
ic_1^2\xi^2\epsilon_{ij}(\theta-c_2^2\eta)=0.
\end{equation}
This requirement leads to the following condition between $\eta$
and $\theta$
\begin{equation}
\label{Eq:c-2}
\eta=c_2^{-2}\theta.
\end{equation}

From Eqs.~(\ref{Eq:hat-a}), (\ref{Eq:c1-c2}) and (\ref{Eq:c-2}) we
obtain the following deformed annihilation and creation operators
$\hat a_i$ and $\hat a_i^\dagger$:
\begin{equation}
\label{Eq:aa+1}
\hat a_i=\sqrt{\frac{1}{2\hbar}\sqrt{\frac{\eta}{\theta}}}\left
(\hat x_i +i\sqrt{\frac{\theta}{\eta}}\hat p_i\right),\;
\hat
a_i^\dagger=\sqrt{\frac{1}{2\hbar}\sqrt{\frac{\eta}{\theta}}}\left
(\hat x_i-i\sqrt{\frac{\theta}{\eta}}\hat p_i\right),
\end{equation}
From Eqs.~(\ref{Eq:xp}) and (\ref{Eq:aa+1}) it follows that the
deformed bosonic algebra of $\hat a_i$ and $\hat a_j^\dagger$
reads \cite{JZZ04a}
\begin{equation}
\label{Eq:[a,a+]1}
[\hat a_i,\hat a_j^\dagger]=\delta_{ij}
+\frac{i}{\hbar}\xi^2\sqrt{\theta\eta}\;\epsilon_{ij},\;
[\hat a_i,\hat a_j]=0,\;(i,j=1,2).
\end{equation}
In Eqs.~(\ref{Eq:[a,a+]1}) the three equations $[\hat a_1,\hat
a_1^\dagger]=[\hat a_2,\hat a_2^\dagger]=1,\;[\hat a_1,\hat
a_2]=0$ are the same as the undeformed bosonic algebra in
commutative space. They constitute a closed algebra. The equation
\begin{equation}
\label{Eq:[a,a+]2}
[\hat a_1,\hat a_2^\dagger] =\frac{i}{\hbar}\xi^2\sqrt{\theta\eta}
\end{equation}
is a new type. Eqs.~(\ref{Eq:[a,a+]1}), including
Eq.~(\ref{Eq:[a,a+]2}), constitute not only a closed but also a
{\it complete} deformed bosonic algebra. Because of
noncommutativity of space, different degrees of freedom are
correlated at the level of the deformed Heisenberg - Weyl algebra
(\ref{Eq:xp}); Eq.~(\ref{Eq:[a,a+]2}) represents such correlations
at the level of the deformed annihilation and creation operators.

The deformed annihilation and creation operators $\hat a_i$ and
$\hat a_i^\dagger$ can be represented by the undeformed ones $a_i$
and $a_i^\dagger$. The general structure of the undeformed
annihilation operator $a_i$ represented by the undeformed phase
space variables $x_i$ and $p_i$ is
$a_i=c_1^{\prime}(x_i +ic_2^{\prime} p_i),$
where the constants $c_1^{\prime}$ can be fixed as follows.
Operators $a_i$ and $a_i^\dagger$ should satisfy bosonic
commutation relations $[a_1,a_1^\dagger]=[a_2,a_2^\dagger]=1$.
From this requirement the undeformed Heisenberg - Weyl algebra
leads to $c_1^{\prime}=\sqrt{1/2\hbar c_2^{\prime}}$. The
undeformed bosonic commutation relation $[a_{i},a_{j}]=0$ is
automatically satisfied, so $c_2^{\prime}$ is a free parameter.
Thus the general structure of the undeformed annihilation and
creation operators reads
\begin{equation}
\label{Eq:aa+2}
a_i=\frac{1}{\sqrt{2\hbar c_2^{\prime}}}(x_i +ic_2^{\prime}p_i),\;
a_i^\dagger=\frac{1}{\sqrt{2\hbar c_2^{\prime}}}(x_i
-ic_2^{\prime}p_i).
\end{equation}
The operators $a_i$ and $a_i^\dagger$ satisfy the undeformed
bosonic algebra
$[a_{i},a_{j}]=[a_i^\dagger,a_j^\dagger]=0, \;
[a_{i},a^{\dagger}_{j}]=i\delta_{ij}$
which constitute a closed and complete algebra in commutative
space.

In the limit $\theta,\eta\to 0$,
the deformed operators $\hat x_{i}, \hat p_{i}, \hat a_{i}, \hat
a_{i}^\dagger$ reduce to the undeformed ones $x_{i}, p_{i}, a_{i},
a_{i}^\dagger$. Eq.~(\ref{Eq:c-2}) indicates that in this limit
$\theta/\eta$ and $\eta/\theta$ should keep finite. It follows
that $c_2=c_2^{\prime}$. From this result and Eqs.~(\ref{Eq:xp}),
(\ref{Eq:hat-x-p}), (\ref{Eq:hat-a})
and (\ref{Eq:aa+2}) it follows
that $\hat a_{i}$ and $\hat a_i^\dagger$ can be represented by
$a_{i}$ and $a_i^\dagger$ as follows:
\begin{equation}
\label{Eq:aa+3}
\hat a_{i}=
\xi(a_{i}+\frac{i}{2\hbar}\sqrt{\theta\eta}\epsilon_{ij}a_{j}),\;
\hat a_{i}^\dagger= \xi(a_{i}^\dagger-
\frac{i}{2\hbar}\sqrt{\theta\eta}\epsilon_{ij}a_{j}^\dagger).
\end{equation}
Similar to Eqs.~(\ref{Eq:xp}) and (\ref{Eq:hat-x-p}), it should be
emphasized that for the case of both position - position and
momentum - momentum noncommuting the scaling factor $\xi$ in
Eq.~(\ref{Eq:aa+3}) guarantees consistency of the framework.
Specially, it maintains the bosonic commutation relation. If one
sets $\xi=1$ in Eq.~(\ref{Eq:aa+3}), it follows that
$[\hat a_{1},\hat a^{\dagger}_{1}]=[\hat a_{2},\hat
a^{\dagger}_{2}]=\left(1+\theta\eta/4\hbar^2\right),$
the bosonic commutation relation $[\hat a_{1},\hat
a^{\dagger}_{1}]=[\hat a_{2},\hat a^{\dagger}_{2}]=1$ cannot be
maintained.

If momentum-momentum is commuting, $\eta= 0$, Eq.~(\ref{Eq:B-E})
shows that the second equation in (\ref{Eq:[a,a+]1}) cannot be
obtained. It is clear that in order to maintain Bose-Einstein
statistics for identical bosons at the non-perturbation level
described by $\hat a_i$ and $\hat a_i^\dagger$ we should consider
both space-space noncommutativity and momentum-momentum
noncommutativity. Eq.~(\ref{Eq:aa+1}) is the most general
representation of the physical annihilation and creation operators
in noncommutative space.
\footnote {\; Related to the general structure of the deformed
annihilation operator Eq.~(\ref{Eq:aa+1}) there is a tacit
understanding about the definition of the physical annihilation
operator such that:

\vspace{0.4cm}

``\; {\it `\,it is possible to construct an infinity of the
creation and annihilation operators which satisfy exactly the
bosonic commutation relations, but do not require any constraint
on the parameters such as Eq.~(\ref{Eq:c-2}).'} For example,
similar to the Landau creation and annihilation operators (acting
within or across Landau levels) involve mixing of spatial
directions in an external magnetic field, we may define the
following annihilation operator
$$
\hat {a_i^{\prime}} = \frac{\nu^{-1}}{\sqrt{2 \hbar c_2^{\prime}}}
\left[ \left( \delta_{ij} - \frac{i c_2^{\prime} \eta}{2 \hbar}
\epsilon_{ij} \right) \hat x_j + i \left(c_2^{\prime} \delta_{ij}
- \frac{i \theta}{2 \hbar} \epsilon_{ij} \right) \hat p_j
\right],$$
where $\nu = \xi (1 - \theta\eta/4 \hbar^2)$. These operators
automatically satisfy the bosonic commutation relations
$[\hat {a_i^{\prime}} , \hat {a_j^{\prime}}^{\dagger} ] =
\delta_{ij},\;
[\hat {a_i^{\prime}} , \hat {a_j^{\prime}}] = [\hat
{a_i^{\prime}}^{\dagger} , \hat {a_j^{\prime}}^{\dagger}]=0. $
Moreover no constraint on the parameters $\theta$ and $\eta$ is
required apart from the obvious one $\eta \theta \ne 4 \hbar^2$.
\;"

\vspace{0.4cm}

Then it follows a related tacit understanding that:

\vspace{0.4cm}

``\; {\it `\,The previous construction also indicates that it is
not compulsory to consider both position and momentum
noncommutativity.'} Indeed, if we take $\eta =0$, $\nu = \xi =1$
in the previous expression for the creation and annihilation
operators, we get:
$$
\hat {a_i^{\prime\prime}} = \frac{1}{\sqrt{2 \hbar c_2'}} \left[
\hat x_i + i \left(c_2' \delta_{ij} - \frac{i \theta}{2 \hbar}
\epsilon_{ij} \right) \hat p_j \right],$$ This is also perfectly
consistent. \;"

\vspace{0.4cm}

In order to clarify the meaning of $\hat {a_i^{\prime}}$ we insert
Eqs.~(\ref{Eq:hat-x-p}) into it. It follows that
$[( \delta_{ij} -i c_2^{\prime} \eta \epsilon_{ij}/2 \hbar) \hat
x_j + i(c_2^{\prime} \delta_{ij} -i \theta \epsilon_{ij}/2 \hbar)
\hat p_j]=\xi (1 - \theta \eta/ 4 \hbar^2)(x_i
+ic_2^{\prime}p_i),$
thus
$$\hat {a_i^{\prime}}=\frac{1}{\sqrt{2 \hbar
c_2^{\prime}}}(x_i +ic_2^{\prime}p_i),$$
which elucidates that $\hat {a_i^{\prime}}$ is just the undeformed
annihilation operator $a_i$ in Eq.~(\ref{Eq:aa+2}), not the
annihilation operator in noncommutative space. This explains that
$\hat {a_i^{\prime}}$ and $\hat {a_i^{\prime}}^{\dagger}$
automatically satisfy the undeformed bosonic commutation
relations, and no constraint on the parameters $\theta$ and $\eta$
is required.

For the case $\eta = 0$, $\nu = \xi =1$, inserting
Eqs.~(\ref{Eq:hat-x-p}) into $\hat {a_i^{\prime\prime}},$ we
obtain
$ \hat {a_i^{\prime\prime}} = (x_i +ic_2^{\prime}p_i)/\sqrt{2
\hbar c_2'}$
which is the annihilation operator in commutative space again.}
Eq.~(\ref{Eq:c-2}) shows that for any system the general feature
of a relation between noncommutative parameters is a direct
proportionality. It is fixed by the consistency of the deformed
Heisenberg - Weyl algebra (\ref{Eq:xp}) with the deformed bosonic
algebra (\ref{Eq:[a,a+]1}).

Normal quantum mechanics in commutative space is a most successful
theory, fully confirmed by experiments. It is correct from the
atomic scale $10^{-10}\; m$ down to at least the scale $10^{-18}\;
m$. It means that any corrections originated from spatial
noncommutativity must be extremely small, thus both noncommutative
parameters $\eta$ and $\theta$ must be extremely small.
\footnote {\; There are different bounds on the parameter $\theta$
set by experiments. The space-space noncommutative theory from
string theory violates Lorentz symmetry and therefore strong
bounds can be placed on the parameter $\theta$, the existing
experiments \cite{CHKLO} give $\theta/(\hbar c)^2\le (10
\;TeV)^{-2}$. Comparing with the above estimation, other bounds on
$\theta$ exist: measurements of the Lamb shift \cite{CST} give a
weaker bound;
clock-comparison experiments \cite{MPR} claim a stronger
bound.
The magnitude of $\theta$ is surely extremely small.}
This is guaranteed by Eq.~(\ref{Eq:c-2}).
\footnote {\; In literature different relations between $\eta$ and
$\theta$ were consideration. For example, Ref.~\cite{Haba}
considered a general D-dimensional case, according their
theoretical framework, the realization of the deformed Heisenberg
- Weyl algebra $[\hat x^{\mu},\hat
x^{\nu}]=i\theta^{\mu\nu},\;[\hat x^{\mu},\hat
p_{\nu}]=i\delta^{\mu}_{\nu},\;[\hat p_{\mu},\hat
p_{\nu}]=i\eta_{\mu\nu}$ with a special relation
$\eta_{\mu\nu}=-(\theta^{-1})_{\mu\nu}$ takes the form: $\hat
x^{\mu}=\frac{1}{2}x^{\mu}-\theta^{\mu\nu}p_{\nu}$ and $\hat
p_{\mu}=p_{\mu}-\frac{1}{2}(\theta^{-1})_{\mu\nu}x^{\nu}$, where
Greek indices $\mu,\nu$ run over $0, 1,\cdots,D-1$. An extremely
small $\theta$ guarantees that corrections originated from spatial
noncommutativity for $\hat x^{\mu}$ is extremely small (In a
correct theoretical framework $\frac{1}{2}x^{\mu}$ should be
$x^{\mu}$ so that $\hat x^{\mu}$ reduces to $x^{\mu}$ in the limit
$\theta\to 0$). But an extremely small $\theta$ corresponds to an
extremely large $\theta^{-1}$. It leads to that corrections from
spatial noncommutativity for $\hat p_{\mu}$ is extremely large.
The case $\eta_{\mu\nu}=-(\theta^{-1})_{\mu\nu}$ does not
correspond to real physics.}

In the context of quantum mechanics how to fix the proportional
coefficient $K$ in Eq.~(\ref{Eq:c-2}) from a first principle is
open.

Now we consider the second aspect. Following the standard
procedure of constructing the Fock space of many - particle
systems at the level of annihilation and creation operators in
commutative space, in the following we take
Eqs.~(\ref{Eq:[a,a+]1}) as the {\it definition relations} for the
complete and closed deformed bosonic algebra without making
further reference to its $\hat x_i$, $\hat p_i$ representations,
generalize it to many - particle systems and find a basis of the
Fock space.

We introduce the following auxiliary operators, the tilde
annihilation and creation operators
\begin{equation}
\label{Eq:tilde-a}
\tilde a_1=\frac{1}{\sqrt{2\alpha_1}} \left(\hat a_1+i\hat
a_2\right),\;
\tilde a_2=\frac{1}{\sqrt{2\alpha_2}} \left(\hat a_1-i\hat
a_2\right),
\end{equation}
where $\alpha_{1,2}=1\pm \xi^2\sqrt{\theta\eta}/\hbar$.
From Eqs.~(\ref{Eq:[a,a+]1}) it follows that the commutation
relations of $\tilde a_i$ and $\tilde a_j^\dagger$ read
\begin{equation}
\label{Eq:tilde[a,a+]}
\left[\tilde a_i,\tilde a_j^\dagger\right]=\delta_{ij},\;
\left[\tilde a_i,\tilde a_j\right]=\left[\tilde a_i^\dagger,\tilde
a_j^\dagger\right]=0,\;(i,j=1,2).
\end{equation}
Thus $\tilde a_i$ and $\tilde a_i^\dagger$ are explained as the
deformed annihilation and creation operators in the tilde system.
The tilde number operators $\tilde N_1=\tilde a_1^\dagger\tilde
a_1$ and $\tilde N_2=\tilde a_2^\dagger\tilde a_2$ commute each
other, $[\tilde N_1,\tilde N_2]= 0.$  A general tilde state
\begin{equation}
\label{Eq:tilde-state}
\widetilde {|m,n\rangle}\equiv (m!n!)^{-1/2}(\tilde
a_1^\dagger)^m(\tilde a_2^\dagger)^n\widetilde {|0,0\rangle},
\end{equation}
where the vacuum state $\widetilde {|0,0\rangle}$ in the tilde
system is defined as $\tilde a_i\widetilde
{|0,0\rangle}=0\;(i=1,2),$ is the common eigenstate of $\tilde
N_1$ and $\tilde N_2$:
$\tilde N_1\widetilde {|m,n\rangle}=m\widetilde {|m,n\rangle}$,
$\tilde N_2\widetilde {|m,n\rangle}=n\widetilde {|m,n\rangle}$,
$(m, n=0, 1, 2,\cdots)$,
and satisfies $\widetilde {\langle m^{\prime},n^{\prime}}
\widetilde {|m,n\rangle}=
\delta_{m^{\prime}m}\delta_{n^{\prime}n}$. Thus $\{\widetilde
{|m,n\rangle}\}$ constitute an orthogonal normalized complete
basis of the tilde Fock space. In the tilde Fock space all
calculations are the same as the case in commutative space, thus
the concept of identical particles is maintained and the formalism
of the deformed Bosonic symmetry which restricts the states under
permutations of identical particles in multi - boson systems can
be similarly developed.

Ref.~\cite{JLR} also investigated the structure of a
noncommutative Fock space and obtained eigenvectors of several
pairs of commuting hermitian operators which can serve as basis
vectors in the noncommutative Fock space. Calculations in such a
noncommutative Fock space are much complex than the above
(commutative) tilde Fock space.

\section{Example}
\setcounter{equation}{0}
In literature
the noncommutative-commutative correspondence has been
investigated and noncanonical  changes of variables in search of
new characteristics have been undertaken \citer{Bell,MM}. In the
tilde system constructed above calculations are easy for systems
whose dynamical behavior can be treated at the level of
annihilation-creation operators, where spatial noncommutativity
are simply represented by parameters $\alpha_i$, not represented
by noncommutative operators.

As usual, harmonic oscillators serve as typical examples. In the
hat system the Hamiltonian of the two-dimensional isotropic
harmonic oscillator reads
\begin{equation}
\label{Eq:H1}
\hat H(\hat x,\hat p)=\frac{1}{2\mu} \hat p_i\hat p_i +
\frac{1}{2}\mu\omega^2 \hat x_i\hat x_i.
\end{equation}
where $\mu$ and $\omega$ are the mass and frequency . In order to
maintain the physical meaning of deformed annihilation-creation
operators $\hat a_i$, $\hat a_i^\dagger$ $(i=1,2)$ the relations
among $(\hat a_i, \hat a_i^\dagger)$ and $(\hat x_i, \hat p_i)$
should keep the same formulation as the ones in commutative space.
Thus they are defined by
\begin{equation}
\label{Eq:aa+2}
\hat a_i=\sqrt{\frac{\mu\omega}{2\hbar}}\left
(\hat x_i +\frac{i}{\mu\omega}\hat p_i\right), \quad \hat
a_i^\dagger=\sqrt{\frac{\mu\omega}{2\hbar}}\left (\hat x_i
-\frac{i}{\mu\omega}\hat p_i\right).
\end{equation}
From the condition $[\hat a_i,\hat a_j]=0$, it follows that
\begin{equation}
\label{Eq:B-E-2}
\eta=\mu^2\omega^2\theta.
\end{equation}
Using Eqs.~(\ref{Eq:aa+2}), the Hamiltonian $\hat H$ is rewritten
by $\hat a_i$ and $\hat a_i^\dagger$ as
\begin{equation}
\label{Eq:H2}
\hat H=\hbar\omega\left (\hat N_1+\hat N_2+1\right),
\end{equation}
where $\hat N_1=\hat a_1^\dagger\hat a_1$ and $\hat N_2=\hat
a_2^\dagger\hat a_2$ are the number operators in the hat system.
Because the
bosonic commutation relation (\ref{Eq:[a,a+]2}) correlates
different degrees of freedom, $\hat N_1$ and $\hat N_2$ do not
commute, $[\hat N_1, \hat N_2]\ne 0.$ They have not common
eigenstates. Though $[\hat a_{i},\hat a_{j}]=[\hat
a_{i}^\dagger,\hat a_{j}^\dagger]=0$,  the deformed Bosonic
symmetry
is not guaranteed in the hat system.
\footnote {\; In the hat system the vacuum state is defined as
$\hat a_i|0,0\rangle=0,\;(i=1,2)$. A general hat state $\widehat
{|m,n\rangle}$ is defined as
\begin{equation*}
\widehat {|m,n\rangle}\equiv c(\hat a_1^\dagger)^m(\hat
a_2^\dagger)^n|0,0\rangle
\end{equation*}
where $c$ is the normalization constant, these hat states
$\widehat {|m,n\rangle}$ are not the eigenstate of $\hat N_1$ and
$\hat N_2$:
\begin{equation*}
\hat N_1\widehat {|m,n\rangle} =m\widehat
{|m,n\rangle}+\frac{i}{\hbar}m\xi^2 \sqrt{\theta\eta}\widehat
{|m+1,n-1\rangle},\; \nonumber
\end{equation*}
\begin{equation*}
\hat N_2\widehat {|m,n\rangle} =n\widehat
{|m,n\rangle}+\frac{i}{\hbar}n\xi^2 \sqrt{\theta\eta}\widehat
{|m-1,n+1\rangle}. \nonumber
\end{equation*}
Because of Eq.~(\ref{Eq:[a,a+]2}), in calculations of the above
equations we should take care of the ordering of $\hat a_i$ and
$\hat a_j^\dagger$ for even $i \ne j$ in the state $\widehat
{|m,n\rangle}$. The states $\widehat {|m,n\rangle}$ are not
orthogonal each other. For example, the inner product between
$\widehat {|1,0\rangle}$ and $\widehat {|0,1\rangle}$ is
\begin{equation*}
\label{Eq:1-2b} \widehat {\langle 1,0|}\widehat {
1,0\rangle}=-\frac{i}{\hbar}\xi^2 \sqrt{\theta\eta}. \nonumber
\end{equation*}
Thus $\{\widehat {|m,n\rangle}\}$ do not constitute an orthogonal
complete basis of the Fock space of a identical - boson system.}
This difficulty can be simply solved in the tilde system. Using
Eqs.~(\ref{Eq:tilde-a}) the Hamiltonian $\hat H$ can be
represented by $\tilde a_i$ and $\tilde a_i^\dagger$ as
\begin{equation}
\label{Eq:tilde-H1}
\hat H=\tilde H=\hbar\omega\left (\alpha_1\tilde
N_1+\alpha_2\tilde N_2+1\right).
\end{equation}
It is worthy noting that {\it all} effects of spatial
noncommutativity are included in the parameters $\alpha_i$, not
represented by noncommutative operators. Because the commutation
relations among $\tilde a_i$, $\tilde a_i^\dagger$ and $\tilde
N_i$ are the same as ones in commutative space, eigenvalues of
$\tilde H$ can be directly read out from Eq.~(\ref{Eq:tilde-H1}),
\begin{equation}
\label{Eq:tilde-E1}
\tilde E_{n_1,n_2} =\hbar\omega\left (\alpha_1 n_1+\alpha_2 n_2+1
\right)=\hbar\omega \left (n_1+n_2+1
\right)+\xi^2\mu\omega^2\theta\left (n_1- n_2 \right),\;(n_1,
n_2=0, 1, 2,\cdots).
\end{equation}
The last term in the above second equation is $\theta$ dependent
which represents the corrections of the energy level originated
from the deformed bosonic algebra (\ref{Eq:[a,a+]1}). There is no
correction for the zero-point energy $\hbar\omega$. It is worth
noting that Eq.~(\ref{Eq:tilde-E1}) gives the {\it exact}
(non-perturbative) eigenvalues.

In order to appreciate noncommutative corrections of the deformed
bosonic algebra (\ref{Eq:[a,a+]1}) to the physical observables, we
compare the above results with ones obtained from the case of only
position - position noncommuting.  For the later case
Ref.~\cite{MM} obtained the energy spectrum of noncommutative
oscillators with mass $\mu$ and frequency $\omega$,
$E_{n_1,n_2}^{\;\prime} =\hbar\Omega\left (n_1+n_2+1
\right)-M\Omega^2\theta\left (n_1- n_2 \right)/2,$
where $1/M\equiv 1/\mu+\mu\theta^2\omega^2/4\hbar^2$ and
$M\Omega^2\equiv \mu\omega^2,$ thus
$\Omega=\left (1+\mu^2\theta^2\omega^2/4\hbar^2\right)^{1/2}\omega
\approx \omega+\mu^2\theta^2\omega^3/8\hbar^2.$
The energy spectrum $E_{n_1,n_2}^{\;\prime}$ can be approximately
represented as
\begin{equation}
\label{Eq:E2}
E_{n_1,n_2}^{\;\prime}\approx \hbar\omega\left (n_1+n_2+1
\right)+\frac{\mu^2\theta^2\omega^3}{8\hbar}\left (n_1+n_2+1
\right) -\frac{1}{2}\mu\omega^2\theta\left (n_1- n_2 \right).
\nonumber
\end{equation}
In the above the $\theta$ dependent terms appreciate the
noncommutative corrections of energy spectrum of noncommutative
oscillators to the commutative ones. It shows that their behavior
of noncommutative corrections is different from ones in
Eq.~(\ref{Eq:tilde-E1}) originated from the deformed bosonic
algebra (\ref{Eq:[a,a+]1}). Specially, there is a shift
$\mu^2\theta^2\omega^3/8\hbar$ of the zero-point energy in
$E_{n_1,n_2}^{\;\prime}$.

\vspace{0.4cm}

\section{Summary and Discussions}
\setcounter{equation}{0}
(i) In the tilde Fock space the deformed Bosonic symmetry is
maintained, the investigation of the consistent deformed bosonic
algebra is completed, all calculations can be similarly developed
as if in commutative space and all effects of spatial
noncommutativity are simply represented by parameters $\alpha_i$.
Such a noncommutative-commutative correspondence in the tilde
system works for general systems whose dynamical behavior can be
investigated at the level of annihilation-creation operators.

(ii) On the fundamental level of quantum field theory the
annihilation and creation operators appear in the expansion of the
(free) field operator $\Psi(x)=\int d^3k a_k(t)\Phi_k(x)+H.c.$
The
consistent multi-particle interpretation requires the usual
(anti)commutation relations among $a_k$ and $a^\dagger_k$.
The noncommutative extension of quantum field theory was obtained
by deforming the ordinary product between quantum fields into the
Moyal "star" product.
For the case of both position - position and momentum - momentum
noncommutativity, however, the corresponding investigation on the
fundamental level of noncommutative quantum field theory is
involved. Noncommutative quantum mechanics, as the one-particle
sector of noncommutative quantum field theory, can be treated in a
more or less self-contained way so that a more detailed study of
quantum systems at the level of noncommutative quantum mechanics
should be useful.
It is expected that some qualitative features obtained at the
level of noncommutative quantum mechanics may survive at the level
of noncommutative quantum field theory. Therefore investigations
of the deformed bosonic algebra at the level of noncommutative
quantum mechanics may give some clue for further development in
noncommutative quantum field theory. Studies on noncommutative
corrections of the deformed bosonic algebra (3.6) on the
fundamental level of quantum field theory will be the next step.


\vspace{0.4cm}

ACKNOWLEDGMENTS

\vspace{0.4cm}

This work has been supported by the Natural Science Foundation of
China under the grant number 10575037 and by the Shanghai
Education Development Foundation.

\clearpage

\end{document}